\documentclass[12pt]{iopart}

\usepackage[pdftex]{graphicx}
\usepackage[breaklinks=true,colorlinks=true,linkcolor=blue,urlcolor=blue,citecolor=blue]{hyperref}
\usepackage[export]{adjustbox}
\usepackage{iopams} 
\usepackage{amstext}
\usepackage{braket}

\usepackage{cite}

\newcommand{\up}{\uparrow}
\newcommand{\down}{\downarrow}

\newcommand{\tx}{\text}

\renewcommand{\vec}[1]{\ensuremath{\mathbf{#1}}}

\newcommand{\Abs}[1]{\ensuremath{\left| #1 \right|}}

\newcommand{\ev}[1]{\langle #1 \rangle}
\newcommand{\vb}[1]{\textbf{#1}}
\newcommand{\eqref}[1]{(\ref{#1})}

\begin{document}

\title[Revealing attractive electron-electron interaction]{Revealing attractive electron-electron interaction in a quantum dot by full counting statistics}
\author{Eric Kleinherbers, Philipp Stegmann, and J\"urgen K\"onig}

\address{Theoretische Physik and CENIDE, Universit\"at Duisburg-Essen, D-47048 Duisburg, Germany}
\ead{eric.kleinherbers@uni-due.de}
\vspace{10pt}
\begin{indented}
\item[]April 2018
\end{indented}

\begin{abstract}
Recent experiments [Nature \textbf{521}, 196 (2015) and Nat. Commun. {\bf 8}, 395 (2017)] have presented evidence for electron pairing in a quantum dot beyond the superconducting regime.
Here, we show that the impact of an \textit{attractive} electron-electron interaction on the \textit{full counting statistics} of electron transfer through a quantum dot is qualitatively different from the case of a repulsive interaction.
In particular, the sign of higher-order (generalized) factorial cumulants reveals more pronounced correlations, which even survive in the limit of fast spin relaxation.
\hphantom{{\it Keywords\/}: }single-electron tunneling, Coulomb blockade
\end{abstract}

\submitto{\NJP}

\maketitle

\section{Introduction}
Attractive electron-electron interaction in BCS superconductors leads to the formation of a condensate of Cooper pairs~\cite{ref:bardeen_theory}. In the context of unconventional and high-$T_c$ superconductors, short-range attractive interactions are considered that generate local electron pairs. Different microscopic origins such as bipolaronic, excitonic, plasmonic, or chemical mechanisms have been proposed~\cite{ref:micnas_superconductivity}. Irrespective of the nature of the interaction, a local attractive pairing potential can be described by a negative-$U$ Anderson impurity model~\cite{ref:anderson_model}. 

Despite the long-running interest, only recently attractive interaction has been studied experimentally on a single electronic orbital
in quantum-dot devices based on carbon nanotubes with auxiliary polarizers~\cite{ref:hamo_repatt} and at a $\text{LaAlO}_3/\text{SrTiO}_3$ interface~\cite{ref:cheng_pairing,ref:cheng_tunable,ref:tomczyk_micrometer,ref:prawiroatmodjo_transport}, where electron-electron attraction has been found even above the critical temperature for superconductivity.
Moreover, theoretical proposals have suggested to utilize the coupling to a mechanical resonator to engineer attractive interaction in a quantum dot~\cite{ref:weiss_spin,ref:szechenyi_electron}.
Experimental evidence for electron-electron attraction has been found in the dependence of the differential conductance on applied bias and gate voltages and magnetic fields.
A very recent theoretical work has explored signatures of the attractive interaction due to a slow driving of parameters~\cite{placke_attractive_2018}.

In this paper, we show that an attractive interaction has a qualitatively different impact on the \textit{full counting statistics}~\cite{ref:lesovik_fcs,ref:levitov_fcs} of electron transfer than a repulsive one.
The electron-transfer statistics is characterized by the probability distribution $P_N(t)$ that $N\geq0$ electrons have been transferred in a given time interval $[0,t]$. It can be measured via a sensitive electrometer such as a quantum point contact~\cite{gustavsson_counting_2005, fujisawa_bidirectional_2006, fricke_bimodal_2007, flindt_universal_2009,rossler_tunable_2013, wagner_optimal_2017} or a single-electron transistor~\cite{lu_real_2003, bylander_current_2005}.

Particularly convenient tools to characterize the distribution~$P_N(t)$ are (generalized) factorial cumulants $C_{s,m}(t)=\partial_z^m \ln{\cal M}_{s}(z,t)|_{z=0}$ of order $m$ derived from the generating function 
\begin{equation}
	{\cal M}_{s}(z,t) := \sum_N (z+s)^N P_N(t)\,.
\end{equation}
Due to the involved $z$-transform, these cumulants are adapted to integer-valued stochastic variables such as the number $N$ of tunneled electrons. 
Choosing the parameter $s$ equal to $1$ yields factorial cumulants.
They can be expressed by a linear combination of factorial moments $\langle N^{(m)}\rangle (t) := \sum_N N^{(m)} P_N(t)$, i.e., expectation values of the falling factorials $N^{(m)}:= N(N-1)\ldots (N-m+1)$. 
The first $C_{1,1}=\langle N\rangle$ and second factorial cumulant $C_{1,2}=\langle N^{2} \rangle- \langle{N}\rangle^2-\langle N \rangle$ are (after dividing by the length $t$ of the time interval) related to the average current and the current noise.
The case $s\neq 1$ yields {\it generalized} factorial cumulants.
The extra factor $s^N$ that is introduced by this generalization allows to study dynamical phase transitions~\cite{garrahan_thermodynamics_2010,carollo_making_2017}, superconducting correlations~\cite{souto_quench_2017}, topologically protected modes~\cite{engelhardt_random_2017, stegmann_inverse_2017}, or deterministic dynamics~\cite{potanina_electron_2017,walldorf_electron_2018,stegmann_coherent_2018}.

Higher-order cumulants contain more information than the average number of transfered charges does.
It is well known that a super-Poissonian Fano factor (equivalent to a positive second-order factorial cumulant $C_{1,2} >0$) indicates interaction-induced correlations between the transferred electrons.
One of the advantages of using factorial cumulants is that its sign can be used as an indicator of correlations not only for the second- but also for higher-order cumulants~\cite{ref:kambly_fcs,ref:kambly_corr,ref:stegmann_gen}.
Furthermore, the possibility to choose the parameter $s$ different from $1$ increases the sensitivity to detect correlations~\cite{ref:stegmann_gen}.
It is, therefore, more than natural to use generalized factorial cumulants to elucidate qualitative differences between the full counting statistics of systems with attractive and repulsive interaction, respectively.

The outline is as follows. First, in section~\ref{sec:model}, we introduce the negative-$U$ Anderson impurity model and present the master-equation formalism to calculate (generalized) factorial cumulants. 
Results are presented in section~\ref{sec:FCS}.
In particular, we find that higher-order factorial cumulants violate a sign criterion for attractive interaction and fulfill it for repulsive interaction.
The requirements on the charge detector to resolve the predicted results are discussed in section~\ref{sec:detector}.
We finish in section~\ref{sec:Conclusions} with the conclusions.
 \begin{figure*}
 {\includegraphics[width=15.54cm]{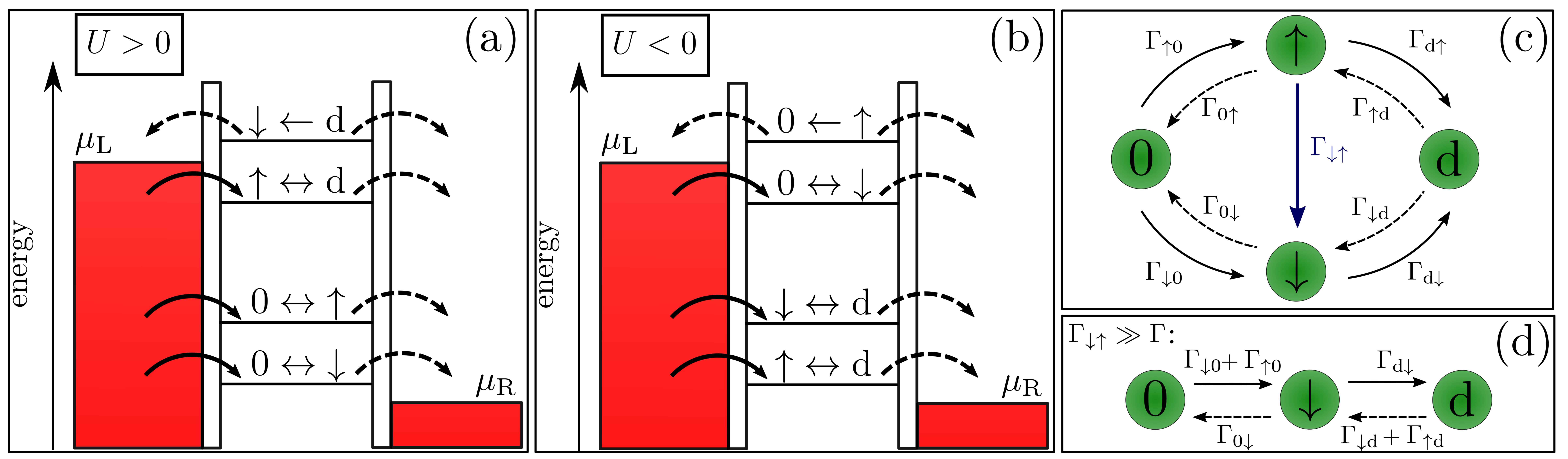}}
\caption{
	(Color online) Excitation energies for the two considered transport scenarios with (a) repulsive ($U>0$) and (b) attractive ($U<0$) electron-electron interaction. The leads are filled with electrons up to the electrochemical potential $\mu_r$ with $r=\text{L},\text{R}$. The excitation energies $E_\chi-E_{\chi^\prime}$ to add or remove one electron are $E_\sigma-E_0=\epsilon_\sigma$ for transitions between $\ket{0}\leftrightarrow\ket{\sigma}$  and $E_\text{d}-E_{\sigma}=\epsilon_{\bar{\sigma}}+U$ for transitions between $\ket{\sigma}\leftrightarrow\ket{\text{d}}$, whereas $\bar{\sigma}$ denotes the opposite spin to $\sigma$.  (c) Quantum-dot states $\ket{\chi}\in\lbrace \ket{0},\ket{\uparrow},\ket{\downarrow},\ket{\text{d}}\rbrace$ and possible transition rates $\Gamma_{\chi\chi^\prime}$ describing a transition from $\ket{\chi^\prime}$ to $\ket{\chi}$. (d) Effective three-state system for a fast spin relaxation $\Gamma_{\downarrow\uparrow}\gg\Gamma$.
	}
\label{scheme}
\end{figure*}
\section{Negative-$U$ Anderson impurity}\label{sec:model}
 The minimal model to study electron transport through a quantum dot subject to attractive interaction is the negative $U$-Anderson impurity model~\cite{ref:koch_pair,ref:koch_neg}
\begin{equation}
\mathcal{H}=\mathcal{H}_\text{dot}+\mathcal{H}_{\text{tun}}+\mathcal{H}_\text{leads}\, .
\end{equation}
The full Hamiltonian is build up from three terms,
\begin{eqnarray}
\label{eq:Hdot}
\mathcal{H}_\text{dot}=\sum_{\sigma=\uparrow,\downarrow} \epsilon_{\sigma} c^\dagger_{\sigma}c_{\sigma}+U c^\dagger_{\uparrow}c_{\uparrow}c^\dagger_{\downarrow}c_{\downarrow}\, ,\\
\mathcal{H}_\text{tun}=\sum_{r \vec k  \sigma }t_{r} c_{\sigma}^{\dagger} a_{r \vec k \sigma} +\text{h.c.}\, , \\
\mathcal{H}_\text{leads}=\sum_{r\vec k \sigma} \varepsilon_{\vec k \sigma} a^{\dagger}_{r \vec k \sigma} a_{r \vec k \sigma}\, .
\end{eqnarray}
The single-level quantum dot is modeled by $\mathcal{H}_\text{dot}$ with fermionic creation (annihilation) operators $c^\dagger_{\sigma}$ ($c_{\sigma}$) for an electron with spin $\sigma=\uparrow,\downarrow$.
Spin degeneracy of the orbital level $\epsilon$ is lifted by the Zeeman splitting $\Delta$ due to an applied magnetic field, $\epsilon_{\uparrow,\downarrow}=\epsilon\pm\Delta/2$. 
The interaction energy $U$ for double occupancy is negative for attractive interaction (and positive for repulsive interaction).
The quantum-dot eigenstates are $\ket{\chi}\in\lbrace \ket{0},\ket{\uparrow},\ket{\downarrow},\ket{\text{d}}\rbrace$ with eigenenergies $E_\chi\in\lbrace 0,\epsilon_\uparrow,\epsilon_\downarrow,\epsilon_\uparrow\text{+}\epsilon_\downarrow\text{+}U \rbrace$ for the empty dot, occupancy with a single spin-$\uparrow$ electron, 
occupancy with a single spin-$\downarrow$ electron, and two electrons on the dot. The resulting excitation energies $E_\chi-E_{\chi^\prime}$ to add or remove one electron are depicted in figure~\ref{scheme}a and b. In particular, for transitions between $\ket{0}\leftrightarrow\ket{\sigma}$ we get $E_\sigma-E_0=\epsilon_\sigma$ and for transitions between $\ket{\sigma}\leftrightarrow\ket{\text{d}}$ the excitation energies are $E_\text{d}-E_{\sigma}=\epsilon_{\bar{\sigma}}+U$, whereas $\bar{\sigma}$ denotes the opposite spin to $\sigma$.

The dot is weakly tunnel coupled to a left $r=\text{L}$ and a right $r=\text{R}$ normal-conducting metallic lead, described by the tunneling Hamiltonian $\mathcal{H}_{\tx{tun}}$. The operators $a^{\dagger}_{r \vec k \sigma}$ ($a_{r \vec k \sigma}$) create (annihilate) an electron with energy $\varepsilon_{\vec k \sigma}$, momentum $\vec k$, and spin $\sigma$ in lead $r$. The metallic leads are modeled as reservoirs of noninteracting electrons $\mathcal{H}_{\text{leads}}$ with flat bands near the Fermi energy and a spin-independent density of states $\rho_r$. A bias voltage $V$ is symmetrically applied such that the electrochemical potentials are $\mu_\text{L}=-\mu_\text{R}=eV/2$.
The time scale of tunneling is set by the tunnel-coupling strength $\Gamma_\text{r}=(2\pi/\hbar) \Abs{t_r}^2 \rho_r$. We emphasize that the tunneling amplitude $t_r$ and the density of states $\rho_r$ enter our calculations only via the parameter $\Gamma_r$. It is convenient to define the sum $2\Gamma=\Gamma_\text{L}+\Gamma_\text{R}$ and the asymmetry $a=(\Gamma_\text{L}-\Gamma_\text{R})/(\Gamma_\text{L}+\Gamma_\text{R})$, implying $\Gamma_\text{L/R}=(1\pm a)\Gamma$. 

Supporting the experimental feasibility of a full counting experiment, which favors a rarely changing dot occupation, we assume a weak tunnel-coupling strength $\hbar \Gamma \ll k_\tx{B}T \ll \vert U\vert,\Delta, eV$ compared to all other energy scales. Therefore, we simulate the dynamics of the system by the $N$-resolved master equation (sketched in figure~\ref{scheme}c)
\begin{eqnarray}
\label{eq:mastereq1}
\dot{P}_N^0=-(\Gamma_{\uparrow 0}{+}\Gamma_{\downarrow 0})P_N^0+\Gamma_{0\uparrow}P_{N-1}^\uparrow(t)+\Gamma_{0\downarrow}P_{N-1}^\downarrow,\\
\dot{P}_N^\uparrow=\Gamma_{\uparrow 0}P_N^0-(\Gamma_{0 \uparrow}{+}\Gamma_{\text{d}\uparrow}{+}\Gamma_{\downarrow\uparrow})P_N^\uparrow+ \Gamma_{\uparrow \text{d}}P_{N-1}^\text{d}, \\
\dot{P}_N^\downarrow=\Gamma_{\downarrow 0}P_N^0+\Gamma_{\downarrow\uparrow}P_N^\uparrow-(\Gamma_{0 \downarrow}{+}\Gamma_{\text{d}\downarrow})P_N^\downarrow + \Gamma_{\downarrow \text{d}}P_{N-1}^\text{d}, \\\label{eq:mastereq4}
\dot{P}_N^\text{d}=\Gamma_{\text{d}\uparrow}P_N^\uparrow +\Gamma_{\text{d}\downarrow}P_N^\downarrow-(\Gamma_{\uparrow \text{d}}{+}\Gamma_{\downarrow \text{d}})P_N^\text{d},
\end{eqnarray}
up to first order in $\Gamma$. Here, $P_N^\chi(t)$ is the probability that $N$ electrons have tunneled out of the quantum dot in a given time interval $[0,t]$ and the dot is in state $\ket{\chi}$ at $t$. Transitions increasing $N$ are depicted by dashed arrows in figure~\ref{scheme}. Opposite transitions (depicted by solid arrows) leave the electron counter $N$ unchanged. 
The sequential tunneling rates $\Gamma_{\chi\chi^\prime}$ describing a transition from $\ket{\chi^\prime}$ to $\ket{\chi}$ are calculated with Fermi's golden rule
\begin{eqnarray}
\label{eq:fermi}
\Gamma_{\sigma 0}=\sum_{r=\text{L},\text{R}} \Gamma_r f_r(\epsilon_\sigma), \\
\Gamma_{0\sigma}=\sum_{r=\text{L},\text{R}} \Gamma_r \lbrack 1-f_r(\epsilon_\sigma)\rbrack, \\
\Gamma_{\text{d}\sigma}=\sum_{r=\text{L},\text{R}} \Gamma_r f_r(\epsilon_{\bar{\sigma}}+U),\\
\Gamma_{\sigma \text{d}}=\sum_{r=\text{L},\text{R}} \Gamma_r \lbrack 1-f_r(\epsilon_{\bar{\sigma}}+U)\rbrack,
\end{eqnarray}
where $f_r(\epsilon)=\lbrace\text{exp}\lbrack(\epsilon-\mu_r)/k_\text{B}T\rbrack+1\rbrace^{-1}$ is the Fermi function of lead $r$. A spin opposite to $\sigma$ is denoted by $\bar{\sigma}$.
Moreover, we also account for spin relaxation by including the spin-flip rate $\Gamma_{\downarrow\uparrow}$ in the master equation~(\ref{eq:mastereq1})-(\ref{eq:mastereq4}). Spin relaxation may originate from hyperfine interaction with a local nuclei bath~\cite{ref:golovach_phonon,ref:khaetskii_nuclei,ref:hanson_overhauser,ref:erlingsson1}.
Spin-orbit interaction on the other hand leads to spin relaxation only for quantum dots with more than one orbital.

For the following, it is convenient to $z$-transform the master equation. We obtain $\dot{\vb{P}}_{\!z}=\vb{W}_{\!z} \vb{P}_{\!z}$ with the probability vector $\vb{P}_{\!z}(t)=\sum_N z^N(P_N^0, P_N^\uparrow, P_N^\downarrow, P_N^\text{d})$ and the generator of the system's dynamics 
\begin{equation}
\label{eq:wz}
	\vb{W}_{\!z}=
	\left(\begin{array}{cccc}
	\text{-}\Gamma_{\uparrow 0}{\text{-}}\Gamma_{\downarrow 0} & z\Gamma_{0\uparrow} & z\Gamma_{0\downarrow} & 0 \\  \Gamma_{\uparrow 0} & \text{-}\Gamma_{0 \uparrow}{\text{-}}\Gamma_{\text{d}\uparrow}{\text{-}}\Gamma_{\downarrow\uparrow} & 0 & z \Gamma_{\uparrow \text{d}} \\  \Gamma_{\downarrow 0} & \Gamma_{\downarrow\uparrow} & \text{-}\Gamma_{0 \downarrow}{\text{-}}\Gamma_{\text{d} \downarrow} & z\Gamma_{\downarrow \text{d}} \\  0 & \Gamma_{\text{d} \uparrow} & \Gamma_{\text{d} \downarrow} & \text{-}\Gamma_{\uparrow \text{d}}{\text{-}}\Gamma_{\downarrow \text{d}}  \\
\end{array}\right).
\end{equation}
The $z$-transformed master equation is solved by $\vb{P}_{\!z}(t)=\text{exp}(\vb{W}_{\!z} t)\vb{P}_{\!\text{st}}$ with the stationary probability distribution $\vb{P}_{\!\text{st}}$ determined by $\vb{W}_{\!1} \vb{P}_{\!\text{st}}=0$ and the normalization $(1,1,1,1)\cdot \vb{P}_{\!\text{st}}=1$. We sum over all states $\chi$ and finally obtain the generating function
\begin{equation}
\label{eq:generfun}
	{\cal{M}}_s(z,t)=(1,1,1,1)\cdot \text{exp}(\vb{W}_{\!z+s} t)\vb{P}_{\!\text{st}}\,
\end{equation}
yielding the generalized factorial cumulants of order $m$

\begin{equation}\label{eq:gencum}
C_{s,m}(t)=\frac{\partial^m {\ln \cal M}_s(z,t)}{\partial z^m} \bigg|_{z=0} \, ,
\end{equation}
or, equivalently $C_{s,m}(t)= \frac{\partial^m {\ln \cal M}_0(z,t)}{\partial z^m} \big|_{z=s}$. Note that the generator of the system's dynamics $\vb{W}_{\!z}$ is proportional to $\Gamma$, see equation \eqref{eq:wz}. Therefore, it follows with equation \eqref{eq:generfun} and \eqref{eq:gencum} that all cumulants $C_{s,m}(\Gamma t)$ are only functions of the product $\Gamma t$. The value of $\Gamma$ is not specified in the following calculations since we choose $\Gamma^{-1}$ as the unit of time.


\section{Revealing interaction by full counting statistics}\label{sec:FCS}
The most general case of uncorrelated electron transfer is that of independent but nonidentical tunneling events described by the Poisson binomial distribution~\cite{ref:wang}
\begin{equation}
\label{eq:Poissbinomial}
	P_N(t) = \sum_{A_N} \prod_{j \in A_N}p_j(t)  \prod_{k \in A_N^\tx{c}} \left[1-p_k(t) \right] \, .
\end{equation}
The distribution is build up from different Bernoulli trials. The sum runs over all sets $A_N$ with $N$ integers selected from the natural numbers $\mathbb{N}$. A set determines $N$ trials succeeding each with a single-electron tunneling probability $p_j(t)$. The complement $A_N^\tx{c}=\mathbb{N}\backslash A_N$ enumerates the not succeeding trials, each with probability $1-p_k(t)$.

For this situation of uncorrelated electron transfer, the generalized factorial cumulants fulfill the sign criterion~\cite{ref:kambly_fcs,ref:stegmann_gen}
\begin{equation}
\label{eq:signc}
	(-1)^{m-1}C_{s,m}(t)\geq0
\end{equation}
for all $m$ if $s\geq0$ and for all even $m$ if $s<0$. 
This criterion is independent of the specific model and transport regime. In particular, it could be used for techniques~\cite{cotunneling_noise,cotunneling_FCS,ridley_ numerically_2018,souto_transient_2018,haertle_decoherence_2013,becker_non_2012} that address transport beyond sequential tunneling.
Any violation of this sign criterion reveals correlated tunneling events. 
Since charge transfer of noninteracting fermions through a two-lead system can always be described as Poisson binomial~\cite{ref:abanov_zeros1}, those correlations must originate from some electron-electron interaction.
As elaborated in the following, attractive interaction leads to a different and much more pronounced violation than repulsive interaction.

We study the transport situation sketched in figure~\ref{scheme}a (for $U>0$) and b (for $U<0$), respectively. The bias voltage $eV$ provides sufficient energy to allow for all quantum-dot states $\ket{0},\ket{\uparrow},\ket{\downarrow}$ and $\ket{\text{d}}$ but is small enough to exclude one of the possible excitation energies. Consequently, for $U>0$ ($U<0$) the transition $\ket{\downarrow}\rightarrow\ket{\text{d}}$ ($\ket{0}\rightarrow\ket{\uparrow}$) is energetically suppressed, see figure \ref{scheme}a and b. The temperature requirements for a sufficient suppression are $k_\text{B}T\ll\epsilon_\uparrow\text{+}U-\mu_\text{L}$ (for $U>0$) and $k_\text{B}T\ll\epsilon_\uparrow-\mu_\text{L}$ (for $U<0$), respectively.
For a larger bias voltage, such that all excitation energies are inside the energy window provided by the electrochemical potentials of the left and right lead, the tunneling rates~(\ref{eq:fermi}) become independent of $U$ and the charge-transfer dynamics is equivalent to that of a noninteracting system, $U=0$.
In the absence of a Zeeman energy it is impossible to exclude only one excitation energy from transport, because the upper two and lower two excitation energies are degenerate, respectively.

\subsection{Fast spin relaxation}
Due to the finite Zeeman energy, spin symmetry is broken, leading to an imbalance between spin-up and spin-down occupation of the quantum dot.
This complicates the dynamics of the stochastic system and correlations are, in general, more likely to occur.
In a real experiment, however, spin relaxation, e.g., due to coupling of the quantum-dot electron to nuclear spins, is often an issue. 
In the limit of fast spin relaxation, as compared to the dwell time of the electrons in the quantum dot, a spin-up electron entering the empty quantum dot immediately relaxes its spin to the energetically favorable spin-down state. 
This simplifies the system's dynamics and one has to check whether and how this affects the correlations in the charge-transfer statistics. 
Therefore, we discuss two limiting cases: first the worst case of fast spin relaxation $\Gamma_{\downarrow\uparrow}\gg \Gamma$ and, then, in the next section the ideal case of slow spin relaxation $\Gamma_{\downarrow\uparrow}\ll \Gamma$.

\begin{figure*}
{\includegraphics[width=8.cm]{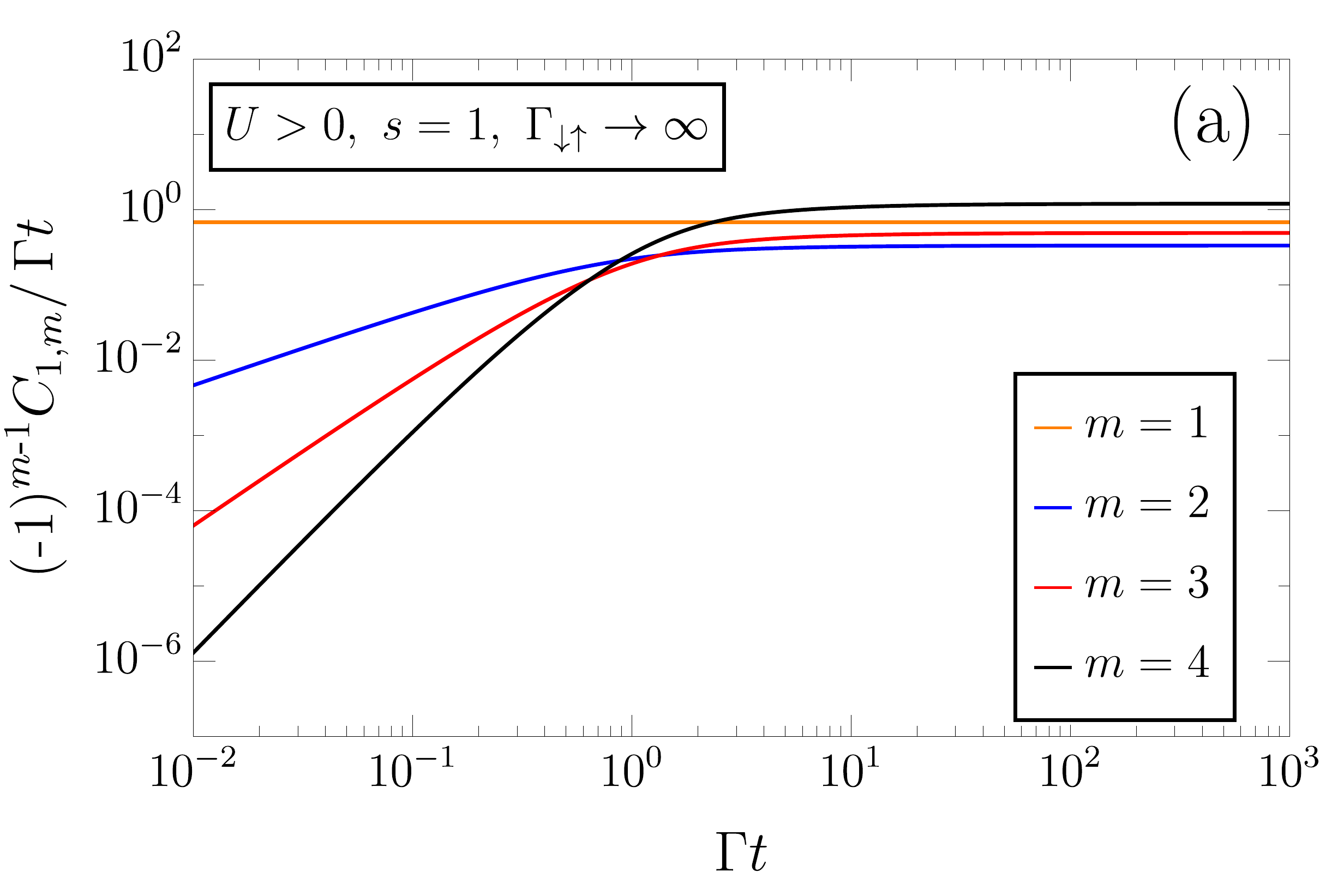}}
{\includegraphics[width=8.cm]{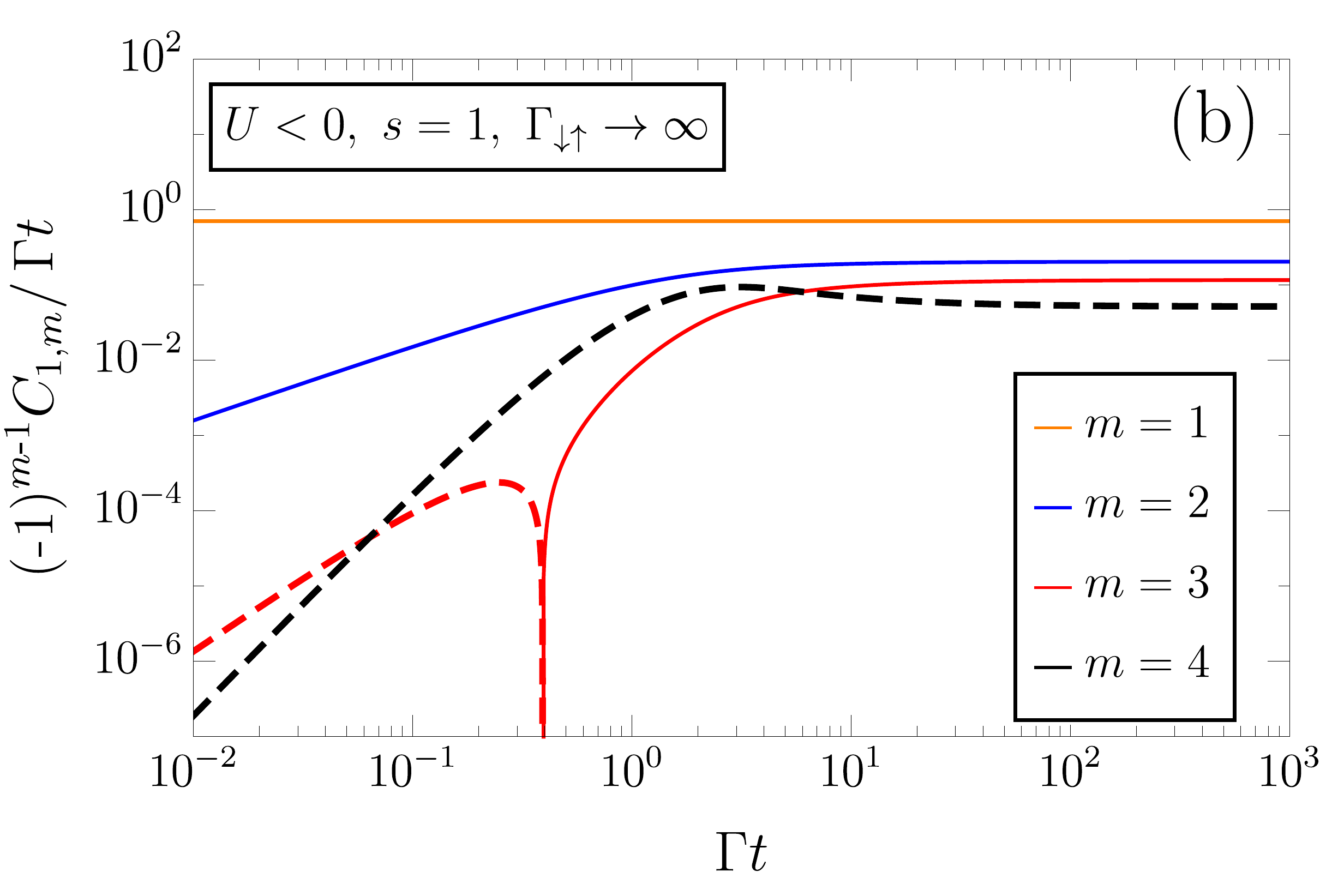}}
\caption{
	(Color online) Factorial cumulants $C_{1,m}(t)$ as function of time $t$ with a fast spin-relaxation rate, $\Gamma_{\downarrow\uparrow}\rightarrow \infty$, for (a) repulsive ($U>0$) and (b) attractive ($U<0$) electron-electron interaction. Other parameters are $\epsilon+U/2= 0.35\,\Abs{U}$, $eV=1.70 \, \Abs{U}$, $\Delta=0.30\, \Abs{U}$, $k_\text{B}T=0.01\,\Abs{U}$, and $a=0.2$. The sign of $(-1)^{m-1}C_{\text{1},m}(t)$ is positive for solid and negative for dashed lines. The latter case indicates correlated electron tunneling due to interaction.  
	}
\label{spin_s1}
\end{figure*}

In figure~\ref{spin_s1}, the factorial cumulants $C_{s,m}(t)$ for $s=1$ are depicted up to the fourth order $m=4$ as a function of length $t$ of the measurement time interval. 
The results for a repulsive interaction ($U>0$) are shown in figure~\ref{spin_s1}a, those for an attractive interaction ($U>0$) in figure~\ref{spin_s1}b.
In order to visualize sign changes in the logarithmic plot, we use solid lines whenever the factorial cumulants obey the sign criterion equation~(\ref{eq:signc}) and dashed lines when the criterion is violated.
As a consequence, any appearance of a dashed line at any time $t$ or for any order $m$ indicates the presence of correlations.

For the first and the second factorial cumulant, the sign criterion is satisfied for both the $U>0$ and the $U<0$ case.
Their values in the long-time limit are related to the electron current~$\langle I \rangle=\lim_{t \to \infty}C_{1,1}(t)/t$ and the zero-frequency noise~$S(0)=\lim_{t \to \infty}2(C_{1,2}+C_{1,1})(t)/t$~\cite{footnote:0}.
Therefore, the Fano factor, that is often used as an indicator of correlations, remains sub-Poissonian.

The advantage of factorial cumulants, however, is that {\it any} order $m$ tests the presence of correlations.
For $U>0$, {\it all} factorial cumulants, including the higher-order ones, fulfill the sign criterion.
This behavior has been recently observed in experiment~\cite{komijani_counting_2013}.
In contrast, for $U<0$, already the third cumulant $C_{\text{1},3}(t)$ displays a violation of the sign criterion for short times $\Gamma t < 0.4$. 
Even stronger, the fourth cumulant $C_{\text{1},4}(t)$ reveals correlated tunneling on all time scales.

The absence of correlations for a repulsive electron-electron interaction (and the presence of correlations for an attractive electron-electron interaction) can be better understood by inspecting~figure~\ref{scheme}c and d. Due to the fast spin relaxation $\Gamma_{\downarrow\uparrow}\gg\Gamma$, transitions from $\ket{\uparrow}$ to either $\ket{\text{d}}$ or $\ket{\text{0}}$ are very unlikely to happen. Before the corresponding tunneling event takes place, the electron spin has already relaxed to $\ket{\downarrow}$. Now, since $\ket{\uparrow}$ is only a short-living intermediate state decaying into $\ket{\downarrow}$ we can exclude it from the dynamics by modifying the ingoing rates for $\ket{\downarrow}$, i.e., $\Gamma_{\downarrow 0}\rightarrow \sum_\sigma \Gamma_{\sigma 0}$ and $\Gamma_{\downarrow \text{d}}\rightarrow \sum_\sigma \Gamma_{\sigma \text{d}}$. Effectively, we mapped the full four-state system onto a three-state system, compare figure~\ref{scheme}c with d. However, for a repulsive electron-electron interaction a transition from $\ket{\downarrow}$ to $\ket{\text{d}}$ is energetically prohibited, see figure~\ref{scheme}a. As a consequence, the probability for a doubly occupied dot is strongly suppressed as well. Thus, for $U>0$ the dynamics can be simplified even further and we arrive at a simple noninteracting model with two states $\ket{0}$ and $\ket{\downarrow}$ only. Such a two-state system cannot exhibit correlations in the charge-transfer statistics.

For an attractive electron-electron interaction, on the other hand, the situation is completely different.
Only the probability to find the dot in state $\ket{\up}$ is strongly suppressed. Beside $\ket{0}$ and $\ket{\downarrow}$, also state $\ket{\text{d}}$ is populated with non-vanishing probability. The system can not be mapped onto a two-state system anymore.

\subsection{Slow spin relaxation}
\begin{figure*}[t]
{\includegraphics[width=8.cm]{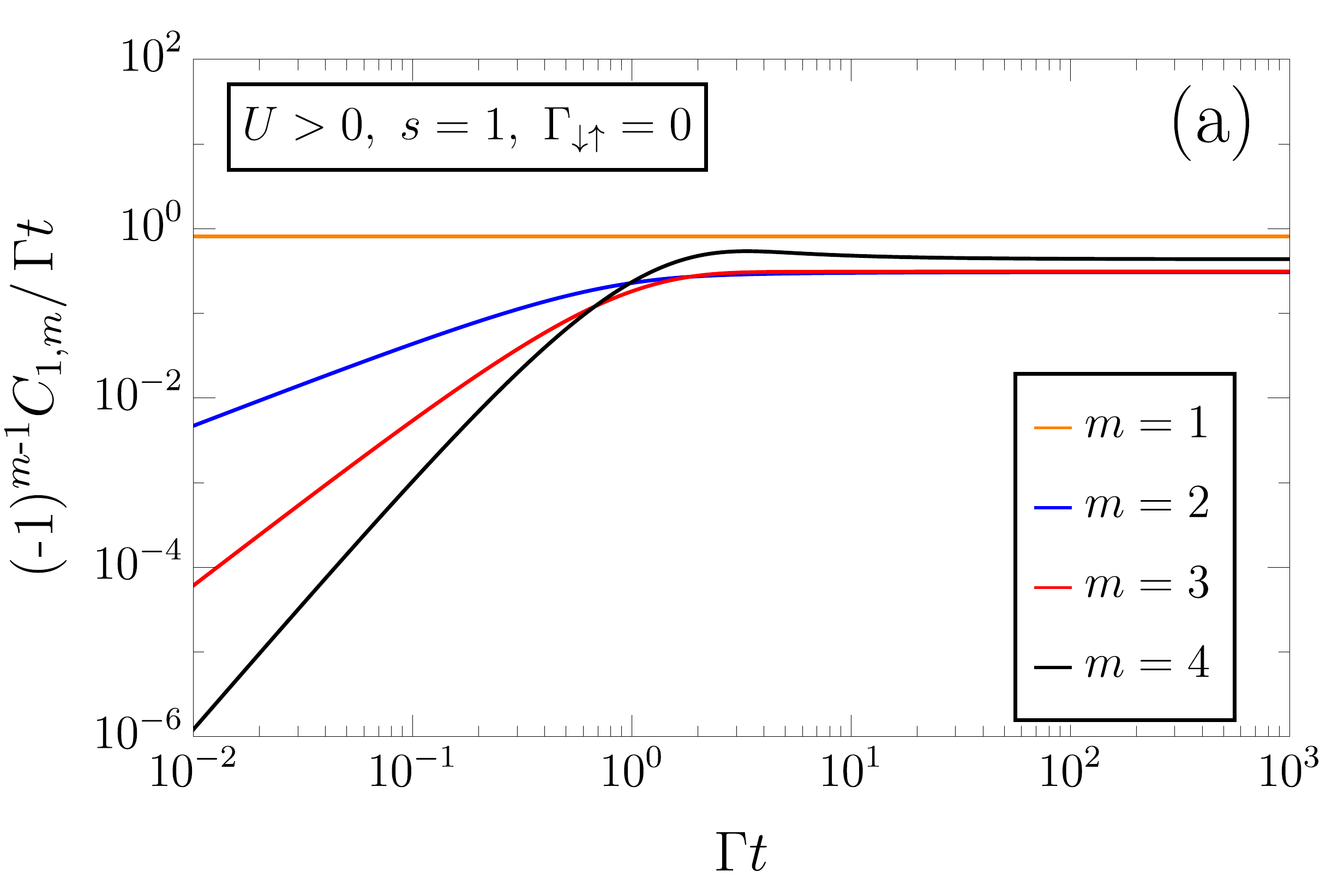}}
{\includegraphics[width=8.cm]{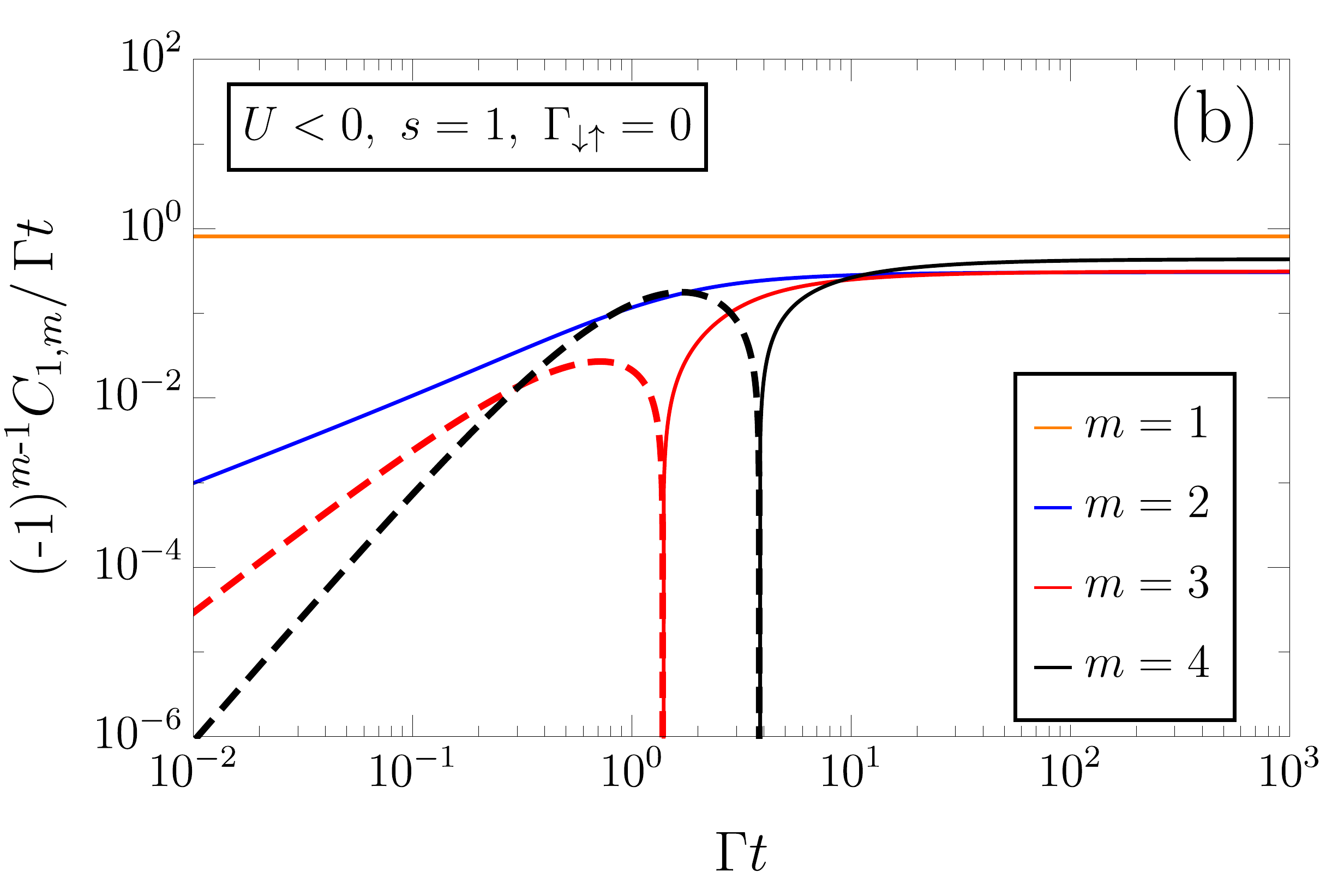}}
\caption{
(Color online) Factorial cumulants $C_{1,m}(t)$ as function of time $t$ with slow spin relaxation, $\Gamma_{\downarrow\uparrow}=0$, for (a) repulsive ($U>0$) and (b) attractive ($U<0$) electron-electron interaction. We chose $a=0$. Other parameters are as in figure~\ref{spin_s1}. The sign of $(-1)^{m-1}C_{1,m}(t)$ is positive for solid and negative for dashed lines. The latter case indicates correlated electron tunneling due to interaction.}
\label{s1}
\end{figure*}

We now address the opposite limit of slow spin relaxation $\Gamma_{\downarrow\uparrow}\ll \Gamma$.
The factorial cumulants ($s=1$) depicted in figure~\ref{s1}a and b show a similar behavior as for the case of fast spin relaxation (see figure~\ref{spin_s1}a and b). 
The first and second factorial cumulants do not reveal any correlation, neither for the $U>0$ nor for the $U<0$ case.
The same is true for the higher-order factorial cumulants  in the case of repulsive interaction.
Attractive interaction, on the other hand, do lead to correlations, indicated by a violation of the sign criterion for the third and fourth factorial cumulant.

\begin{figure*}[t]
{\includegraphics[width=8.cm]{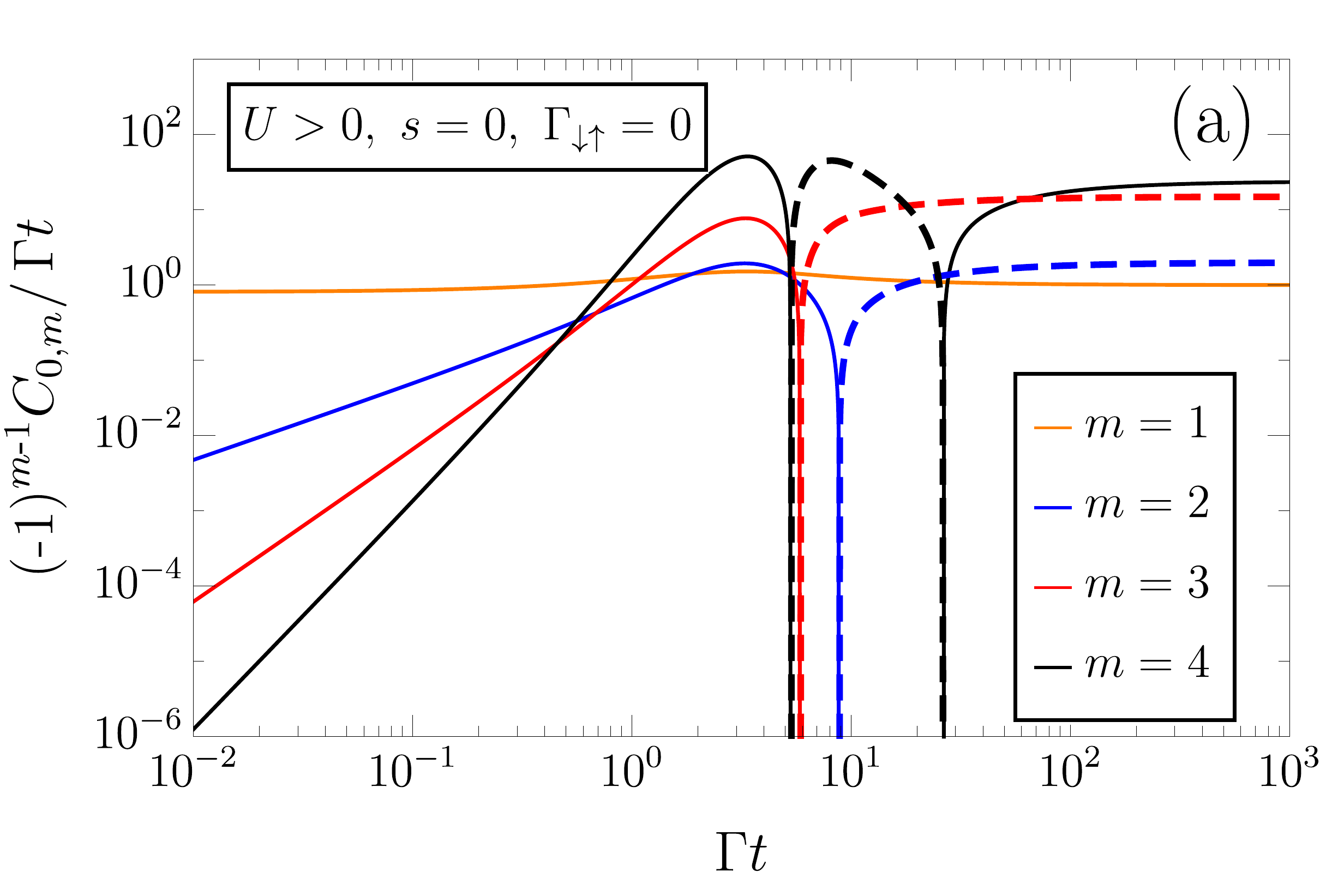}}
{\includegraphics[width=8.cm]{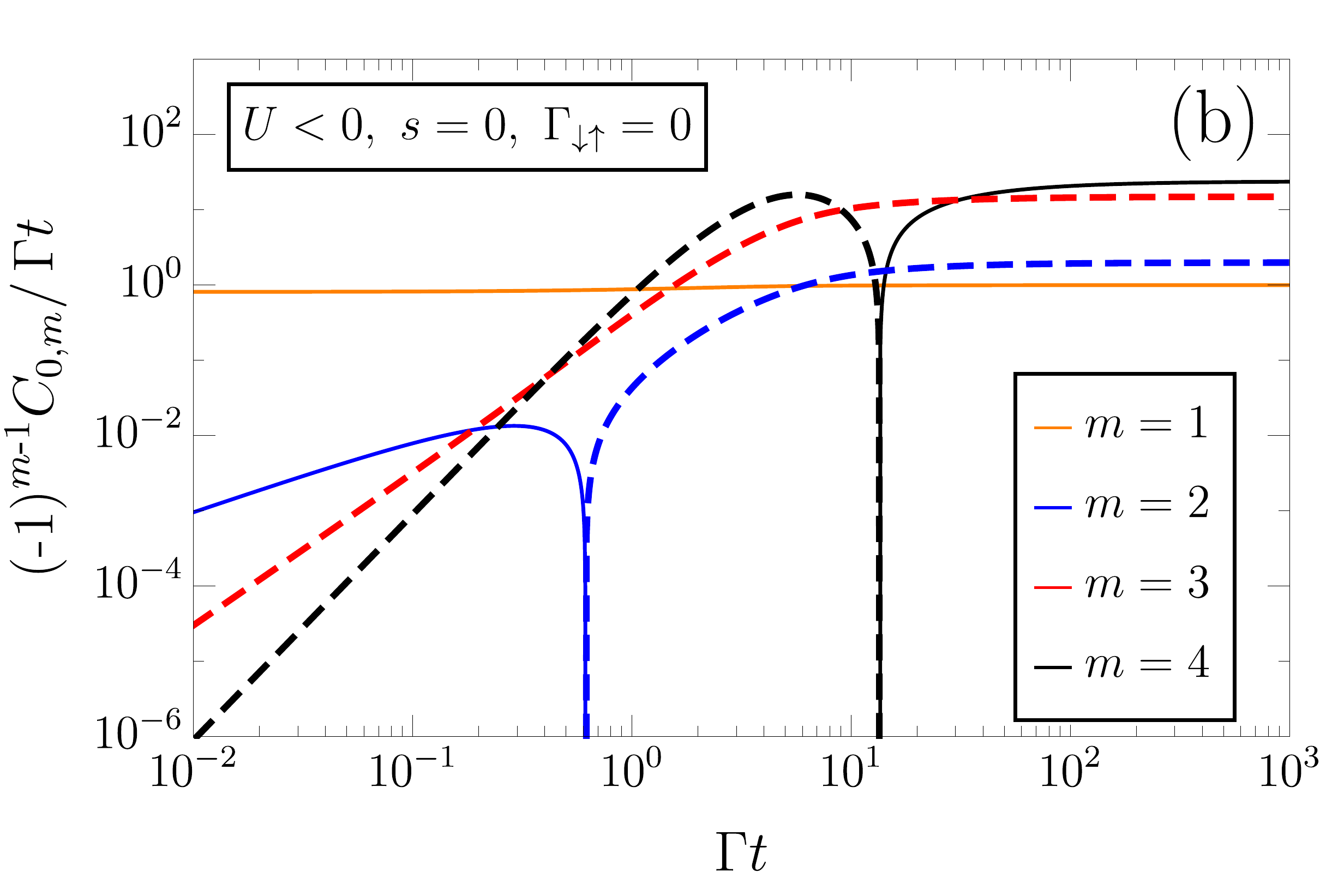}}
\caption{
(Color online) Generalized factorial cumulants $C_{0,m}(t)$ with $s=0$ as function of time $t$ with slow spin relaxation, $\Gamma_{\downarrow\uparrow}=0$, for (a) repulsive ($U>0$) and (b) attractive ($U<0$) electron-electron interaction. We chose $a=0$. Other parameters are as in figure~\ref{spin_s1}. The sign of $(-1)^{m-1}C_{0,m}(t)$ is positive for solid and negative for dashed lines. The latter case indicates correlated electron tunneling due to interaction.}
\label{s0}
\end{figure*}
The absence of any spin relaxation increases the possibility for correlations. 
In particular, the mapping to a two-state system for $U>0$ does not work anymore since, now, also the doubly-occupied state $\ket{\text{d}}$ can occur with finite probability. Before state $\ket{\uparrow}$ relaxes, a spin-$\down$ electron can tunnel into the quantum dot leading to doubly occupancy.
Nevertheless, the qualitative behavior of the factorial cumulants is insensitive to the presence or absence of spin relaxation.  

To increase the sensitivity for correlations, we shift the parameter $s$ of the generalized factorial cumulants $C_{s,m}(t)$ from $s=1$ to smaller values~\cite{ref:stegmann_gen}.
As an example, we choose $s=0$.
Then, we indeed observe a violation of the sign criterion both for repulsive (figure~\ref{s0}a) and attractive interaction (figure~\ref{s0}b) in the long-time limit. 
Furthermore, we find that with the reduced $s$, already the second-order cumulant is sufficient to detect correlations, at least in the long-time limit.

The behavior of the cumulants in the short-time limit, on the other hand, seems to be hardly affected by the parameter $s$: the repulsive case does not display any correlations of any order $m$, while the attractive case shows a violation of the sign criterion for higher-order cumulants.
To understand the short-time behavior in more detail, we expand the cumulants in dynamical Lee-Yang zeros~\cite{brandner_experimental_2017} of the factorial generating function~\cite{ref:stegmann_shorttime}.  
In leading order in time, we can approximate ${\cal{M}}_{1}(z,t)\approx 1+ zP_1(t)+ z^2P_2(t)$ with $P_1(t)\propto t$ and $P_2(t)\propto t^2$. 
Terms including $P_{N\ge 3}\propto t^{2N-2}$ decay too fast with decreasing time to be of relevance for the zeros.
We obtain the zeros $z_{1,2}(t)=-2/\Big\lbrack P_1(t) \pm \sqrt{P_1^2(t)-4P_2(t)} \Big\rbrack\propto t^{-1}$. Inserting those two zeros into the expansion
\begin{equation}
(-1)^{m-1}C_{s,m}(t)=(m-1)!\sum_{j}\frac{1}{z_j^{m}}\propto t^m \label{eq:shorttimecum}
\end{equation}
yields analytic expressions for the generalized factorial cumulants with short-time behavior $t^m$. 
Furthermore, a violation of the sign criterion can only be achieved for $4P_2(t)>P_1^2(t)$. Whereas a change between attractive and repulsive interaction does not (for asymmetry $a\neq$ 0 only slightly) alter $P_1(t)$, the effect on the probability $P_2(t)$ is considerably strong. We obtain in the short-time limit
\begin{equation}\label{eq:P2short}
P_2(t)= \frac{1}{2}\left(\Gamma_{0\uparrow} \Gamma_{\uparrow \text{d}} +\Gamma_{0 \downarrow} \Gamma_{\downarrow \text{d}} \right) P^\text{d}_\text{st}t^2 +\mathcal{O}(t^3) \,.
\end{equation}
For repulsive electron-electron interaction, the stationary probability $P^\text{d}_\text{st}$ to find the system in the state of double occupation is significantly lower than for attractive interaction, compare figure~\ref{scheme}a and b. 
Hence, $P_2(t)$ plays a less important role for $U>0$ than for $U<0$.

In the repulsive case, the electrons tend to avoid each other on short time scales. The transfer statistics can, therefore, be described by independent single-electron tunneling events, i.e., by equation~(\ref{eq:Poissbinomial}). In the case of attractive interaction the probability for two successive sequential tunneling-out events is too high to be compatible with independent electron tunneling.

We emphasize once more that this paper considers the limit of small tunnel-coupling strength
$\hbar \Gamma \ll k_\text{B}T\ll\vert U\vert,\Delta,eV$. Therefore, we account only for sequential tunneling events and neglect the rarely occurring tunneling events of higher-order in $\Gamma$ (e.g., pair tunneling).

\section{Finite-resolution detector}\label{sec:detector}
\begin{figure*}
{\includegraphics[width=8.cm]{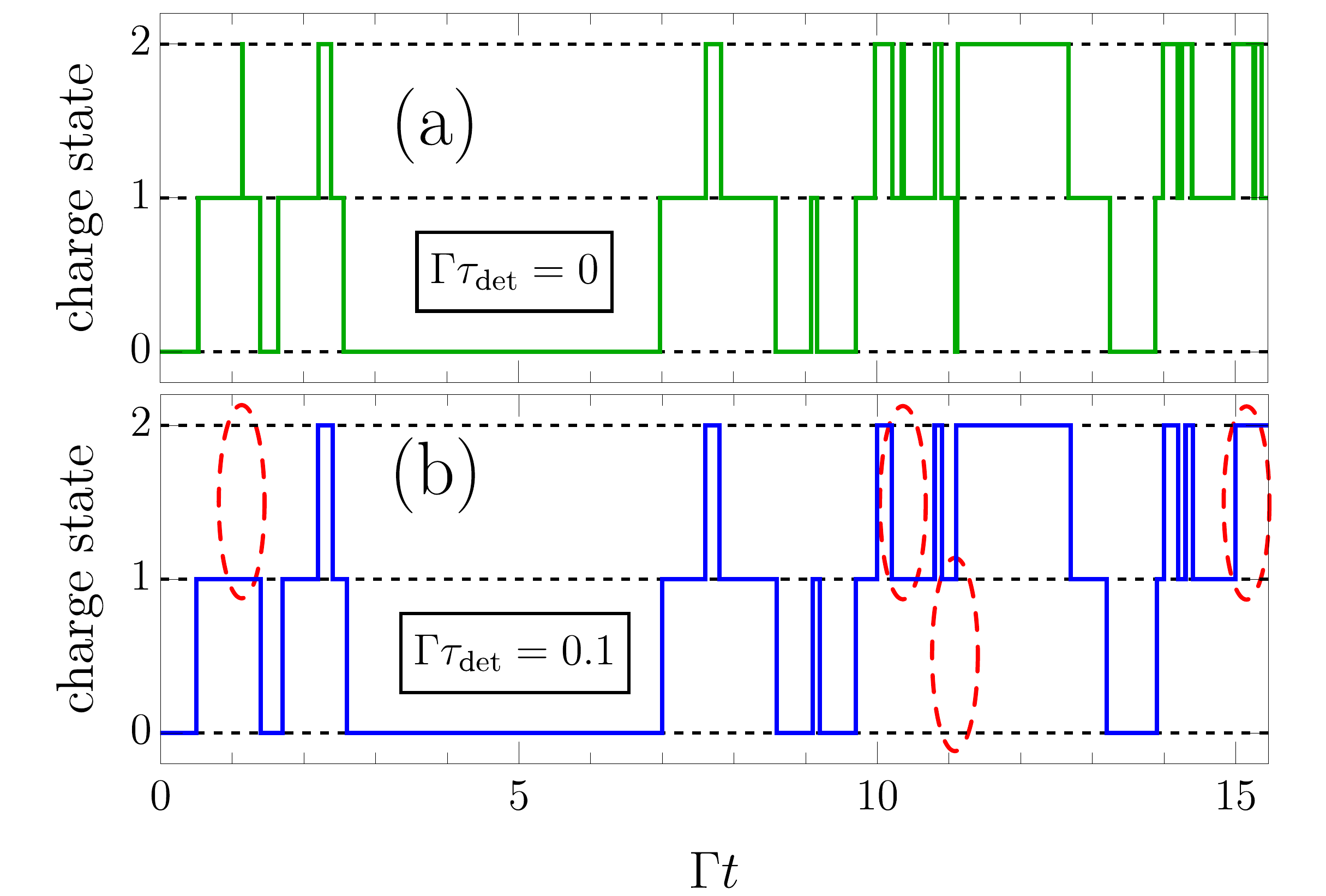}}
{\includegraphics[width=8.cm]{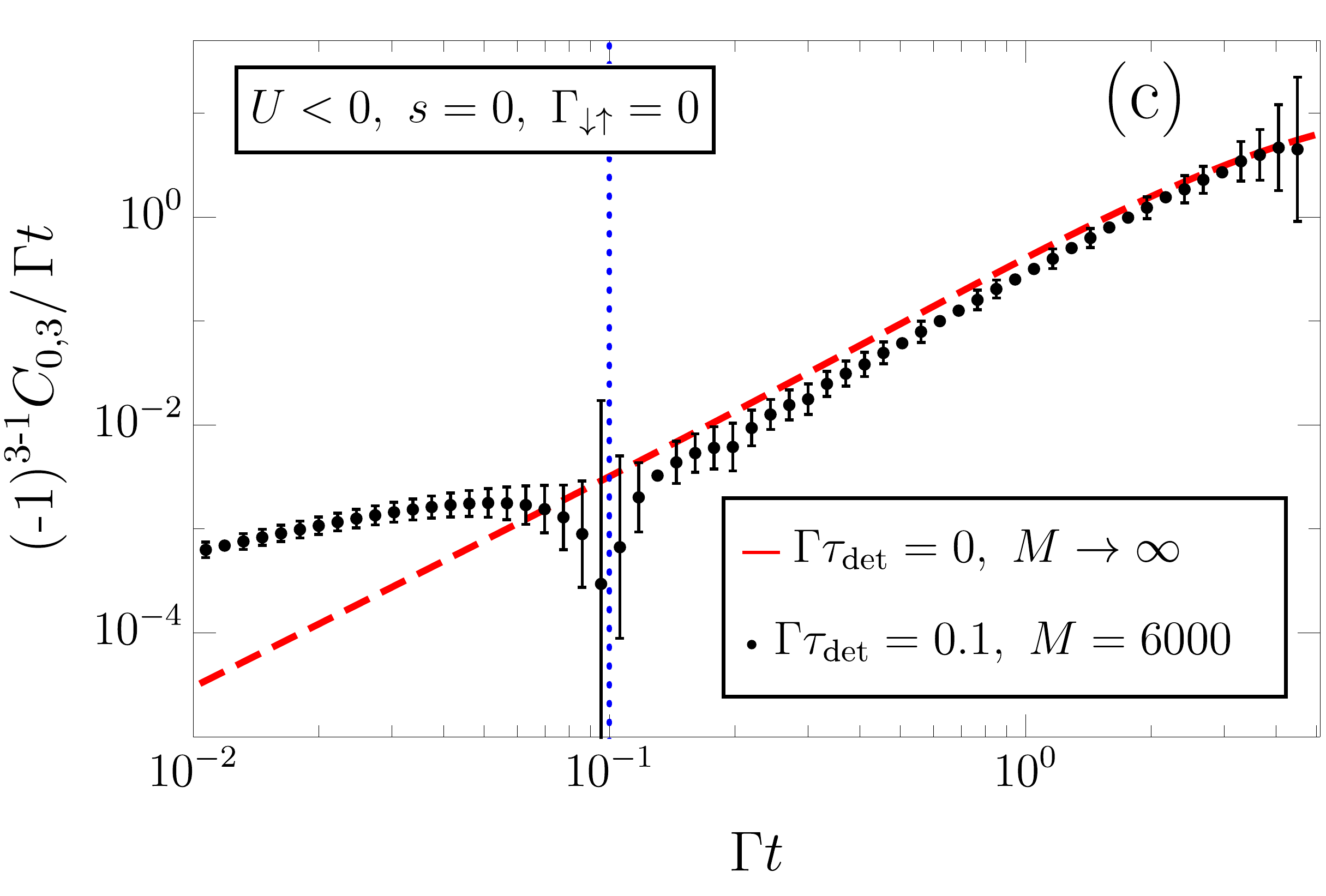}}
\caption{
	(Color online) Example of a simulated time trace for $U<0$ with $M=35$ tunneling events for (a) perfect detector resolution $\tau_\text{det}=0$ and (b) finite detector resolution $\Gamma\tau_\text{det}=0.1$. In (b), missing tunneling events are highlighted with red dashed lines. (c)
Generalized factorial cumulant $C_{0,3}(t)$ with $s=0$ as function of time $t$ with slow spin relaxation, $\Gamma_{\downarrow\uparrow}=0$, for an attractive ($U<0$) electron-electron interaction. We chose $a=0$. Other parameters are as in figure~\ref{spin_s1}. The numerical results are calculated from an ensemble of $50$ time traces, each with $M=6000$ tunneling events and a sampling time $\Gamma\tau_\text{det}=0.1$. The black dots give the mean value $(-1)^{m-1}\ev{C_{0,3}}_{50}$ and the error bars represent the standard deviation $\delta C_{0,3}$. The analytical results $(-1)^{m-1}C_{0,3}$ are shown as a dashed red line. The dotted vertical line marks the sampling time $\Gamma\tau_\text{det}=0.1$. 
	}
\label{det}
\end{figure*}
Both a finite amount of data and a finite detector resolution limit the accuracy of any results deduced from experimental data.
Therefore, we estimate in this last part of the paper whether the discussed differences between $U>0$ and $U<0$ can be confirmed in an actual experiment despite the aforementioned limitations.
We simulate a stepwise time trace that records the number of electrons $\lbrace 0,1,2 \rbrace$ on the quantum dot at every instant of time, see figure~\ref{det}a. We start with some initial state $\ket{\chi}$ and then, weighted by the probability distribution $h_\chi(\tau_{\text{dwell}})=\sum_{\chi^\prime}\Gamma_{\chi^\prime \chi} e^{-\sum_{\chi^\prime}\Gamma_{\chi^\prime \chi} \tau_{\text{dwell}}}$, the system resides for the dwell time $\tau_{\text{dwell}}$ in that state $\ket{\chi}$. Afterwards, weighted by the conditional probabilities $\text{Pr}({\chi^\prime\vert\chi})=\Gamma_{\chi^\prime\chi}/(\sum_{\chi^\prime}\Gamma_{\chi^\prime\chi})$, the system jumps into state $\ket{\chi^\prime}$ given it is in state $\ket{\chi}$.
By repeating this procedure $M$ times, we simulate a time trace that can be interpreted as a measurement signal of an ideal detector with perfect resolution.
In a next step, we artificially incorporate a finite resolution $\tau_\text{det}$ of the detector, which is also referred to as sampling time of the detector. Therefore, we discretize the continuous signal in intervals of length $\tau_\text{det}$ and calculate the average electron number (rounded to $\lbrace 0,1,2 \rbrace$) for each interval. As a consequence, details in the time trace happening on short-time scales $t<\tau_\text{det}$ are averaged out, compare figure~\ref{det}b to figure~\ref{det}a. In particular, tunneling events are slightly shifted in time by less than $\tau_\text{det}$ and some are discarded (highlighted with a red dashed line in figure \ref{det}b).

In figure~\ref{det}c, we illustrate the influence of a finite amount of data ($M$ tunneling events) and the finite sampling time ($\tau_\text{det}>0$).
The red line is the exact analytic result for $C_{0,3}(t)$ from figure~\ref{s0}b, which is recovered for $M\to \infty$ and $\tau_\text{det}\to 0$.
The black dots $(-1)^{m-1}\langle{C_{0,3}(t)}\rangle_{50}$ are obtained from an ensemble average over $50$ cumulants simulated by $50$ different time traces (each with $M=6000$ tunneling events and $\Gamma \tau_\text{det}=0.1$). The error bars represent the standard deviation $\delta C_{s,m}(t)$ from that ensemble average.

We observe that with a moderate detector resolution $\Gamma \tau_\text{det}=0.1$ and a relatively short time trace $M=6000$, the exact result can already be reproduced quite well for times $\Gamma \tau_\text{det} < \Gamma t < 4$, including the sign and the time dependence $C_{0,3} \propto t^3$ [see equation \eqref{eq:shorttimecum}].

The finite amount of data $M$ causes the increasing error bars for $\Gamma t > 4$.
The finite sampling time $\tau_\text{det}$ leads to the slightly underestimated absolute value of $C_{0,3}(t)$ for $ \Gamma\tau_\text{det} < \Gamma t < 4$, because the detector is blind to a certain fraction of tunneling events. The missing number of events can be estimated from the difference between exact and simulated first factorial cumulant $C_{1,1}=\langle{N}\rangle$ (not depicted here).
A more drastic effect of the finite sampling time is the dip near $ t = \tau_\text{det}$ --- highlighted with a vertical blue dotted line --- and the strong deviations for $t<\tau_\text{det}$. Due to the finite sampling time, the detector registers direct transitions between $\ket{0}$ and $\ket{\text{d}}$. Those artificial two-electron processes (pair-tunneling events) result in completely different statistics for short times $t\ll \tau_\text{det}$. Instead of equation \eqref{eq:shorttimecum}, we observe in the short-time limit for all even cumulants $C_{s,m}^{\text{even}}\propto t^{m/2}$ and for all odd cumulants $C_{s,m}^{\text{odd}}\propto t^{(m+1)/2}$~\cite{ref:stegmann_shorttime}.
As a consequence, despite a small variance in the experimental data, the cumulants should not be trusted below $t<\tau_\text{det}$.

\section{Conclusions}\label{sec:Conclusions}
To conclude, we have illustrated that repulsive and attractive interaction in a quantum dot lead to qualitatively different full counting statistics of electron transfer.
Using the sign of (generalized) factorial cumulants of any order $m$ as an indicator of the presence of correlations, we find that an attractive electron-electron interaction has a much stronger tendency to generate correlations than a repulsive one. 
This finding is robust against spin-relaxation that are often present in real samples. 
To reveal the qualitatively different behavior for repulsive and attractive interaction, factorial cumulants of order three or higher need to be tested; the first and second-order factorial cumulants, related to the average current and the current noise, are not sufficient.
By making use of the parameter $s$ of the generalized factorial cumulants, the sensitivity to correlations can be enhanced such that the second-order cumulant already displays a violation of the sign criterion.
The short-time limit is particularly suited to highlight the difference between the $U>0$ and the $U<0$ case.
The robust correlations for $U<0$ in the short-time limit can be attributed to two successive tunneling-out events occurring with a strong enhancement of the probability for double occupancy that is absent for $U>0$.

To the best of our knowledge, there is so far in the literature no report of a measurement of a factorial cumulant that violates the sign criterion, which would be a non-debatable signature and direct proof of correlations in the charge-transfer statistics.  
We propose that a quantum dot with attractive electron-electron interaction is an ideal model system to go for it since correlations are more pronounced and more robust against spin relaxation than for a repulsive interaction. 
Experimental setups reported in Refs.~\cite{ref:cheng_pairing,ref:cheng_tunable,ref:prawiroatmodjo_transport}, extended by a sensitive electrometer such as a quantum point contact or a single-electron transistor, seem perfectly suited to test our proposal.
We estimate that already a moderate amount of data, i.e., $M=6000$ tunneling events, and a moderate detector resolution of $\Gamma\tau_\text{det}=0.1$ is sufficient to reveal the effect of the attractive interaction on the charge-transfer statistics.

\ack
We acknowledge financial support of the Deutsche Forschungsgemeinschaft (DFG) under project KO $1987/5-2$ and SFB 1242 (project A02).


\section*{References}
\begin{thebibliography}{50}
\bibitem{ref:bardeen_theory}
Bardeen~J, Cooper~L~N and Schrieffer~J~R 1957
Theory of Superconductivity
\textit{Phys. Rev.} {\bf 108} 1175
\bibitem{ref:micnas_superconductivity} Micnas~R, Ranninger~J and Robaszkiewicz~S 1990
Superconductivity in narrow- band systems with local nonretarded attractive interactions
\textit{Rev. Mod. Phys.} {\bf 62} 113 
\bibitem{ref:anderson_model} Anderson~P~W 1975
Model for the electronic structure of amorphous semiconductors
\textit{Phys. Rev. Lett.} {\bf 34} 953 
\bibitem{ref:hamo_repatt} Hamo~A, Benyamini~A, Shapir~I, Khivrich~I, Waissman~J, Kassbjerg~K, Oreg~Y, von~Oppen~F and Ilani~S 2016
Electron attraction mediated by Coulomb repulsion
\textit{Nature (London)} {\bf 535} 395
\bibitem{ref:cheng_pairing} Cheng~G, Tomczyk~M, Lu~S, Veazey~J~P, Huang~M, Irvin~P, Ryu~S, Lee~H, Eom~C-B, Hellberg~C~S and Levy~J 2015
Electron pairing without superconductivity
\textit{Nature (London)} {\bf 521} 196
\bibitem{ref:cheng_tunable} Cheng~G, Tomczyk~M, Tacla~B, Lee~H, Lu~S, Veazey~J~P, Huang~M, Irvin~P, Ryu~S,  Eom~C-B, Daley~A, Pekker~D and Levy~J 2016
Tunable Electron-Electron Interactions in ${\mathrm{LaAlO}}_{3}/{\mathrm{SrTiO}}_{3}$ Nanostructures
\textit{Phys. Rev. X} {\bf 6} 041042
\bibitem{ref:tomczyk_micrometer}Tomczyk~M, Cheng~G, Lee~H, Lu~S, Annadi~A, Veazey~J~P, Huang~M, Irvin~P, Ryu~S, Eom~C-B and Levy~J 2016
Micrometer-Scale Ballistic Transport of Electron Pairs in ${\mathrm{LaAlO}}_{3}/{\mathrm{SrTiO}}_{3}$ Nanowires
\textit{Phys. Rev. Lett.} {\bf 117} 096801
\bibitem{ref:prawiroatmodjo_transport} Prawiroatmodjo~G~E~D~K, Leijnse~M, Trier~F, Chen~Y, Christensen~D~V, von~Soosten~M, Pryds~N and Jespersen~T~S 2017
Transport and excitations in a negative-U quantum dot at the ${\mathrm{LaAlO}}_{3}/{\mathrm{SrTiO}}_{3}$ interface
\textit{Nat. Commun.} {\bf 8} 395
\bibitem{ref:weiss_spin} Weiss~S, Br\"uggemann~J and Thorwart~M 2015
Spin vibronics in interacting nonmagnetic molecular nanojunctions
\textit{Phys. Rev. B} {\bf 92} 045431 
\bibitem{ref:szechenyi_electron} Sz\'echenyi~G, P\'alyi~A and Droth~M 2017
Electron-electron attraction in an engineered electromechanical system
\textit{Phys. Rev. B} {\bf 96} 245302
\bibitem{placke_attractive_2018} Placke~B, Pluecker~T, Splettstoesser~J and Wegewijs~M~R 2018
Attractive and driven interaction in quantum dots: mechanisms for geometric pumping
arXiv:1803.08479
\bibitem{ref:lesovik_fcs} Levitov~L~S and Lesovik~G~B 1993
Charge distribution in quantum shot noise
\textit{JETP Lett.} {\bf 58} 230 
\bibitem{ref:levitov_fcs} Levitov~L~S, Lee~H and Lesovik~G~B 1996
Electron counting statistics and coherent states of electric current
\textit{Journal of Mathematical Physics} {\bf 37} 4845 
\bibitem{gustavsson_counting_2005} Gustavsson~S, Leturcq~R, Simovi\v{c}~B, Schleser~R, Ihn~T, Studerus~P, Ensslin~K, Driscoll~D~C and Gossard~A~C 2006
Counting Statistics of Single Electron Transport in a Quantum Dot
\textit{Phys. Rev. Lett.} {\bf 96} 076605 
\bibitem{fujisawa_bidirectional_2006} Fujisawa~T, Hayashi~T, Tomita~R and Hirayama~Y 2006
Bidirectional Counting of Single Electrons
\textit{Science} {\bf 312} 1634
\bibitem{fricke_bimodal_2007}
Fricke~C, Hohls~F, Wegscheider~W and Haug~R~J 2007
Bimodal counting statistics in single-electron tunneling through a quantum dot
\textit{Phys. Rev. B} {\bf 76} 155307
\bibitem{flindt_universal_2009} Flindt~C, Fricke~C, Hohls~F, Novotn{\'y}~T, Neto{\v c}n{\'y}~K, Brandes~T and Haug~R~J 2009
Universal oscillations in counting statistics
\textit{PNAS} {\bf 106} 10116
\bibitem{rossler_tunable_2013} R\"ossler~C, Kr\"ahenmann~T, Baer~S, Ihn~T, Ensslin~K, Reichl~C and Wegscheider~W 2013
Tunable charge detectors for semiconductor quantum circuits
\textit{New J. Phys.} {\bf 15} 033011
\bibitem{wagner_optimal_2017} Wagner~T, Bayer~J~C, Rugeramigabo~E~P and Haug~R~J 2017
Optimal single-electron feedback control
\textit{Phys. Status Solidi B} {\bf 254} 1600701
\bibitem{lu_real_2003}
Lu~W, Ji~Z, Pfeiffer~L, West~K~W and Rimberg~A~J 2003
Real-time detection of electron tunneling in a quantum dot 
\textit{Nature (London)} {\bf 423} 422
\bibitem{bylander_current_2005}
Bylander~J, Duty~T and Delsing~P 2005
Current measurement by real-time counting of single electrons
\textit{Nature (London)} {\bf 434} 361
\bibitem{garrahan_thermodynamics_2010} Garrahan~J~P and Lesanovsky~I 2010
Thermodynamics of Quantum Jump Trajectories
\textit{Phys. Rev. Lett.} {\bf 104} 160601 
\bibitem{carollo_making_2017} Carollo~F, Garrahan~J~P, Lesanovsky~I and P\'erez-Espigares~C 2018
Making rare events typical in Markovian open quantum systems
\textit{Phys. Rev. A} {\bf 98} 010103(R)
\bibitem{souto_quench_2017} Souto~R~S, Mart\'{\i}n-Rodero~A and Yeyati~A~L 2017
Quench dynamics in superconducting nanojunctions: Metastability and dynamical Yang-Lee zeros
\textit{Phys. Rev. B} {\bf 96} 165444 
\bibitem{engelhardt_random_2017} Engelhardt~G, Benito~M, Platero~G, Schaller~G and Brandes~T 2017
Random-walk topological transition revealed via electron counting
\textit{Phys. Rev. B} {\bf 96} 241404(R) 
\bibitem{stegmann_inverse_2017} Stegmann~P and K\"onig~J 2017
Inverse counting statistics based on generalized factorial cumulants
\textit{New J. Phys.} {\bf 19} 023018
\bibitem{potanina_electron_2017} Potanina~E and Flindt~C 2017
Electron waiting times of a periodically driven single-electron turnstile
\textit{Phys. Rev. B} {\bf 96} 045420
\bibitem{walldorf_electron_2018} Walldorf~N, Padurariu~C, Jauho~A-P and Flindt~C 2018
Electron Waiting Times of a Cooper Pair Splitter
\textit{Phys. Rev. Lett.} {\bf 120} 087701
\bibitem{stegmann_coherent_2018} Stegmann~P, K\"onig~J and Weiss~S 2018
Coherent dynamics in stochastic systems revealed by full counting statistics
\textit{Phys. Rev. B} {\bf 98} 035409
\bibitem{ref:kambly_fcs} Kambly~D and Flindt~C 2013
Time-dependent factorial cumulants in interacting nano-scale systems
\textit{J. Comput. Electron.} {\bf 12} 331
\bibitem{ref:kambly_corr} Kambly~D, Flindt~C and B{\"u}ttiker~M 2011
Factorial cumulants reveal interactions in counting statistics
\textit{Phys. Rev. B} {\bf 83} 075432
\bibitem{ref:stegmann_gen} Stegmann~P, Sothmann~B, Hucht~A and K\"onig~J 2015
Detection of interactions via generalized factorial cumulants in systems in and out of equilibrium
\textit{Phys. Rev. B} {\bf 92} 155413
\bibitem{ref:koch_pair} Koch~J, Raikh~M~E and von~Oppen~F 2006
Pair Tunneling through Single Molecules
\textit{Phys. Rev. Lett.} {\bf 96} 056803
\bibitem{ref:koch_neg} Koch~J, Sela~E, Oreg~Y and von~Oppen~F 2007
Nonequilibrium charge-Kondo transport through negative-$U$ molecules
\textit{Phys. Rev. B} {\bf 75} 195402 (2007).
\bibitem{ref:golovach_phonon} Golovach~V~N, Khaetskii~A and Loss~D 2004
Phonon-Induced Decay of the Electron Spin in Quantum Dots
\textit{Phys. Rev. Lett.} {\bf 93} 016601 
\bibitem{ref:khaetskii_nuclei} Khaetskii~A, Loss~D and Glazman~L 2003
Electron spin evolution induced by interaction with nuclei in a quantum dot
\textit{Phys. Rev. B} {\bf 67} 195329
\bibitem{ref:hanson_overhauser} Hanson~R, Kouwenhoven~L~P, Petta~J~R, Tarucha~S and Vandersypen~L~M~K 2007
Spins in few-electron quantum dots
\textit{Rev. Mod. Phys.} {\bf 79} 1217
\bibitem{ref:erlingsson1} Erlingsson~S~I and Nazarov~Y~V 2002
Hyperfine-mediated transitions between a Zeeman split doublet in GaAs quantum dots: The role of the internal field
\textit{Phys. Rev. B} {\bf 66} 155327
\bibitem{ref:wang} Wang~Y~H 1993
On the number of successes in independent trials
\textit{Stat. Sin.} {\bf 3} 295
\bibitem{cotunneling_noise} Thielmann~A, Hettler~M~H, K\"onig~J and Sch\"on~G 2005
Cotunneling Current and Shot Noise in Quantum Dots
\textit{Phys. Rev. Lett.} {\bf 95} 146806 
\bibitem{cotunneling_FCS} Braggio~A, K\"onig~J and Fazio~R 2006
Full Counting Statistics in Strongly Interacting Systems: Non-Markovian Effects
\textit{Phys. Rev. Lett.} {\bf 96} 026805
\bibitem{ridley_ numerically_2018} Ridley~M, Singh~V~N, Gull~E and Cohen~G 2018
Numerically exact full counting statistics of the nonequilibrium Anderson impurity model
\textit{Phys. Rev. B} {\bf 97} 115109
\bibitem{souto_transient_2018} Souto~R~S, Avriller~R, Yeyati~A~L and Mart\'{\i}n-Rodero~A 2018
Transient dynamics in interacting nanojunctions within self-consistent perturbation theory
arXiv:1803.07372
\bibitem{haertle_decoherence_2013} H\"artle~R, Cohen~G, Reichman~D~R and Millis~A~J 2013
Decoherence and lead-induced interdot coupling in nonequilibrium electron transport through interacting quantum dots: A hierarchical quantum master equation approach
\textit{Phys. Rev. B} {\bf 88} 235426
\bibitem{becker_non_2012} Becker~D, Weiss~S, Thorwart~M and Pfannkuche~D 2012
Non-equilibrium quantum dynamics of the magnetic Anderson model
\textit{New J. Phys.} {\bf 14} 073049
\bibitem{ref:abanov_zeros1} Abanov~A~G and Ivanov~D~A 2009
Factorization of quantum charge transport for noninteracting fermions
\textit{Phys. Rev. B} {\bf 79} 205315
\bibitem{footnote:0}
Technically, to obtain the quantities $\langle{I}\rangle$ and $S(0)$ concordant with a conventional current noise measurement, we have to count directional, i.e., electrons tunneling in opposite directions are counted with opposite signs. However, this leads only to slight differences.
\bibitem{komijani_counting_2013} Komijani~Y, Choi~T, Nichele~F, Ensslin~K, Ihn~T, Reuter~D and Wieck~A~D 2013
Counting statistics of hole transfer in a $p$-type GaAs quantum dot with dense excitation spectrum
\textit{Phys. Rev. B} {\bf 88} 035417
\bibitem{brandner_experimental_2017} Brandner~K, Maisi~V~F, Pekola~J~P, Garrahan~J~P and Flindt~C 2017
Experimental Determination of Dynamical Lee-Yang Zeros
\textit{Phys. Rev. Lett.} {\bf 118} 180601
\bibitem{ref:stegmann_shorttime} Stegmann~P and K\"onig~J 2016
Short-time counting statistics of charge transfer in Coulomb-blockade systems
\textit{Phys. Rev. B} {\bf 94} 125433
\end{thebibliography}
\end{document}